# Highly sensitive and selective sugar detection by terahertz nano-antennas

*Dong-Kyu Lee[1, 2], Ji-Hun Kang[3], Jun-Seok Lee[4], Hyo-Seok Kim[1, 5], Chulki Kim[1], Jae Hun Kim[1], Taikjin Lee[1], Joo-Hiuk Son[2], Q-Han Park[6], and Minah Seo[1†]*

[1]Sensor System Research Center, Korea Institute of Science and Technology (KIST), Seoul 136-791, Republic of Korea

[2]Department of Physics, University of Seoul, Republic of Korea

[3]Department of Physics, University of California at Berkeley, Berkeley, California 94720, USA

[4]Molecular Recognition Research Center, Korea Institute of Science and Technology (KIST), Seoul 136-791, Republic of Korea

[5]Department of Electronics and Communications Engineering, Kwang-woon University, Seoul, Republic of Korea

[6]Department of Physics, Korea University, Seoul 136-701, Republic of Korea

[†] To whom correspondence should be addressed. [†]mseo@kist.re.kr

ABSTRACT

Molecular recognition and discrimination of carbohydrates are important because carbohydrates perform essential roles in most living organisms for energy metabolism[1] and cell-to-cell communication[2]. Nevertheless, it is difficult to identify or distinguish various carbohydrate molecules owing to the lack of a significant distinction in the physical or chemical characteristics[3, 4]. Although there has been considerable effort to develop a sensing platform for individual carbohydrates selectively using chemical receptors or an ensemble array, their detection and discrimination limits have been as high in the millimolar concentration range[5, 6]. Here we show a highly sensitive and selective detection method for the discrimination of carbohydrate molecules using nano-slot-antenna array-based sensing chips which operate in the terahertz (THz) frequency range (0.5–2.5 THz). This THz metamaterial sensing tool recognizes various types of carbohydrate molecules over a wide range of molecular concentrations. Strongly localized and enhanced terahertz transmission by nano-antennas can effectively increase the molecular absorption cross sections, thereby enabling the detection of these molecules even at low concentrations. We verified the performance of nano-antenna sensing chip by both THz spectra and images of transmittance. Screening and identification of various carbohydrates can be applied to test even real market beverages with a high sensitivity and selectivity.

There has been growing interest in the label-free detection of molecules and their analysis by optical detection system including plasmonic methodologies[7], resonant microcavities[8], optical-fibre sensors[9], and interferometry-based biosensors[10]. Because an optics-based label-free sensing system can provide quantitative counting and avoid any complexity from fluorescence tagging, terahertz (THz) time-domain spectroscopy (TDS) for molecular detection is also of greatly increasing importance[11, 12]. The THz TDS is basically a non-contact, non-destructive, and label-free sensing tool suitable for the examination of biological[13] and chemical substances[14], especially, for the direct demonstration of intermolecular signatures within the broad THz spectrum[15]. Recently, metamaterial-based THz sensing platforms were introduced in order to increase the detection sensitivity of small molecules or microorganisms[16]. In particular, subwavelength THz metamaterials on the order of $\lambda/10$–$\lambda/10,000$ can induce a huge field enhancement in transmission[17]. Then, the locally enhanced THz field can be readily used to detect chemical and biological substances with a high sensitivity, because the molecular absorption cross sections are effectively increased by the field enhancement[18].

Here, we present a nano-slot-antenna array-based THz sensing method supporting the highly accurate measurement of even very small quantities of sugar molecules and the selective identification of different sugar molecules. Our designed nano-antenna allows detecting molecules over a very wide concentration range from hundreds of micromoles to tens of moles. Furthermore, the imaging results for THz transmittance demonstrate the remarkable selectivity working for only the targeted sugar molecule. Finally, we demonstrate the detection of sugar levels for real market beverages including diet sodas known to contain a very small amount of artificial sweeteners, which can lead to a new type of non-contact and non-invasive sensing application including a precise sugar monitoring.

We performed THz TDS of various sugar samples in pellet forms. A commercial THz TDS system (Zomega THz Z-3XL) was used to obtain THz spectra in the frequency range of 0.5–2.5 THz (Methods). We measured the THz spectra for several carbohydrate pellets; for D-glucose (a monosaccharide which has the simplest form of a carbohydrate, Fig. 1(a)), fructose (a monosaccharide, Fig. 1(b)), sucrose (a disaccharide which is a combination of the monosaccharides D-glucose and fructose, Fig. 1(c)), and cellulose (a polysaccharide which consists of a linear chain of many linked D-glucose units, Fig. 1(d)). And we plotted the absorption coefficients, $\alpha = \frac{4\pi\kappa}{\lambda} = \frac{4\pi\kappa}{c}f$, where $c$ is the speed of light, and $f$ is the frequency. The hydroxyl functional group has strong absorption peaks, as shown in the figure, e.g. at 1.4 THz, 1.7 THz, and 1.8 THz for D-glucose (Fig. 1(a)), fructose (Fig. 1(b)), and sucrose (Fig. 1(c)), respectively. In stark contrast, the polysaccharide group has no recognizable absorption features in the THz spectrum owing to the complexity of the chains between various structural groups and the lack of coordinated hydrogen-bond vibrations in the crystalline state (Fig. 1(d))[19].

We suggest a novel type of plot-colour contour mapping of the THz absorption coefficients in terms of the frequency which helps us easily recognize various types of carbohydrates in accordance with their spectral characteristics. A monosaccharide group was mapped using a red tone (mannose, galactose, D-glucose, arabinose, and fructose), a disaccharide group was mapped using a yellow tone (maltose and sucrose), a polysaccharide group was mapped using a green tone (cellulose, glycogen, and amylose), and artificial sweeteners were mapped using a blue tone (aspartame and acesulfame K). We took notice of several pronounced absorption peak positions in the spectra in Fig. 1(e) (white and black dashed circles) and then designed the nano-slot antennas for highly sensitive and selective THz sensing, even after dissolution with a very low concentration of molecules.

A schematic of the nano-antenna experiment is shown in Fig. 2(a). Our designed nano-slot-antenna array can induce strong THz field localization and a very large field enhancement in transmission[17], thereby effectively increasing the absorption cross section of sugar molecules[18]. We designed two different nano-slot-antenna arrays for specific sugars: the glucose antenna has a length of $l = 40$ μm (the targeted frequency, $f_{res} = 1.4$ THz, is for D-glucose absorption) and the fructose antenna has a length of $l = 35$ μm (the targeted frequency, $f_{res} = 1.7$ THz, is for fructose absorption). Several interesting sugars including D-glucose, sucrose, and cellulose were measured with the glucose antenna at first.

We progressively changed the molecular concentrations of the sugars from 0 to 500 mg/dL (0–27.5 mmol/L), which seems quite reasonable because the concentration of the normal fasting glucose level in blood is 70–100 mg/dL (3.9 to 5.5 mmol/L), whereas that for patients with diabetic symptoms is 100–125 mg/dL (5.6 to 6.9 mmol/L) and above[20]. The transmitted THz spectrum for D-glucose molecules at a concentration of 250 mg/dL on bare Si shows no distinguishable features compared to the spectrum for a bare Si wafer (Fig. 2(b)) because of the extremely small absorption cross section of the molecule at the reliable frequency regime. The nano-antenna with 1.4 THz resonance was applied to detect D-glucose molecules with concentrations varying from 0 to 4168 mg/dL (Fig. 2(c)). A strongly localized and enhanced THz field by nano-antenna resulted in a significant increase in the absorption cross section, making the D-glucose molecules clearly visible. The estimated THz field enhancement here is approximately 50, as found in an earlier work[21]. It is noted that the concentration for this experiment varies from 10 to 4168 mg/dL, which completely covers three orders of magnitude of concentration levels.

We applied the glucose antenna to two other carbohydrates: sucrose and cellulose with same concentration range. The most drastic change in transmittance was observed in the D-glucose measurement among the three samples because the D-glucose molecule has a clear absorption peak at 1.4 THz, whereas the others do not. A small change in the transmittance for sucrose (Fig. 2(d)) is responsible for the small feature at 1.4 THz in the absorption spectrum (Fig. 1(c)), and a rare change in transmittance is observed for the cellulose sample. The maximum values of the normalized transmittance, $T_{max}$, for D-glucose, sucrose, and cellulose are plotted in terms of the concentration levels with exponential decay fittings (Fig. 2(e)), which have clear sample-dependent decay constants. The decreased transmittances for different samples exactly reflect the absolute values of the absorption coefficients of the molecules in Fig. 1, showing the high performance of our system. The resonance frequency is shifted toward a lower frequency with the value $\Delta f$ as the molecular concentration increases (Fig. 2(f)) with a strong sample dependence as well.

Further measurements with the nano-antenna array targeted to fructose with the resonance frequency at 1.7 THz (fructose antenna) represent a variety of selections according to specific sugar molecules. A notable drastic suppression in transmittance was observed for the fructose molecule; however, a lower change was measured for the D-glucose molecule (Fig. 3(a)). This verifies that our antenna, specifically designed for a certain sugar molecule, excellently works for only the targeted molecule which has a strong absorption at that frequency, but it is insensitive to other molecules. The performance of the sugar antenna with molecular concentration and sample dependencies is now described with finite-difference time-domain (FDTD) calculations.

Transmittance spectra were calculated for two different samples: each has an absorption peak at 1.7 THz and 1.4 THz, respectively (Methods). It is clearly shown that the maximum values in transmittances for the sample possessing an absorption resonance at 1.7 THz is distinctively stronger than that for the other sample possessing an absorption resonance at 1.4

THz (Fig. 3(b)). From the good agreement between FDTD calculations and experimental results, we propose a simple model to explain the molecular-selective detection. The direct transmission of THz light can experience an exponential decay in amplitude while propagating within the lossy cladding medium. Therefore, the change in the peak transmittance can be written as $\frac{T_{sam}}{T_{ref}} = Ae^{-\kappa k h}$, where $T_{ref}$ and $T_{sam}$ are the maximum transmittances through the nano-antenna without and with the cladding; $A$ is the transmittance ratio at the air–cladding boundary; $\kappa$ is the imaginary part of the refractive index of the cladding; $k = 2\pi/\lambda$ is the incidence momentum; and $h$ is the cladding thickness, proportional to the molecular concentration. It is discussed that suppression of the peak transmittance strongly occurs when the molecule possesses high absorption at the resonance frequency of the nano-antenna and predicts the suppression behaviour with a thick cladding well.

The high sensitivity and selectivity of our sugar antenna can be also verified by THz far-field imaging in transmittance (Methods). The fructose antenna was used for the complete discrimination of fructose from D-glucose. The white dashed square lines denote the total nano-antenna area, and the magenta and blue lines represent the dropped stains of fructose (upper-right corner) and D-glucose (lower-left corner) solutions, respectively (Fig. 3(c)). The THz transmittance image at 1.7 THz (Fig. 3(d)) clearly shows different colours between the two sample areas and the slot-antenna-pattern area (middle) owing to different absorptions by the sugar molecules, promising high selectivity for two different sugars. A greater change in colour was observed for the fructose area than the glucose area, as the fructose antenna targeted to.

The fructose antenna was used to detect the sugars contained in various popular sweetened beverages including Coca-Cola Classic, Pepsi-Cola, and Sprite, and in particular, some diet sodas with very low concentrations of sweeteners (Coca-Cola Light and Coca-Cola Zero).

The results for the differently decreased transmittances for various beverages showed the clear existence of the sugar content with different concentrations (Fig. 4(a)). We compared the change in the transmittance with the known nutritional values from the manufacturer and a previous report (Fig. 4(b)); for example, Coca-Cola Classic has a sugar content of 10420 mg/dL in total (fructose 6252 mg/dL and 4168 mg/dL glucose)[22]. Because the nano-antenna array is the most sensitive at a concentration level in the range of tens to hundreds of milligrams per decilitre, we specifically focused on the type of diet sodas which have extremely low concentrations of sweeteners such as aspartame and acesulfame K, e. g. tens of mg/dL. The measured large decreases in the transmittances for the two diet sodas are caused by two sweeteners with high absorption features in the range of 1.7–1.79 THz[23] (Fig. 4(c)). It is notable that our sensing level is valid for monitoring a very small amount of sweetener, even in real market beverages, because these artificial sweeteners have recently received considerable attention owing to their possible addictiveness and toxicity.

In conclusion, nano-antennas operating in the broad THz frequency region can provide highly sensitive and selective detection of carbohydrate molecules, even at concentrations of a few hundred micromoles. Our sugar-antenna-based THz sensing chip exhibits complete selectivity for the targeted sugar molecules, which has the possibility for further applications such as non-invasive blood-sugar monitoring. Further, THz imaging using the sugar antenna with two different samples at once strongly demonstrates the high sensitivity and selectivity of the sugar antennas. Finally, the high performance of the sugar antenna was shown with the real market beverages.

**Methods**

**1. THz time-domain spectroscopy and imaging system**

A commercial THz TDS system (Zomega Z-3XL) based on a Ti:sapphire femtosecond laser with a repetition rate of 80 MHz and a wavelength centred at 800 nm is used. The THz-wave source module consisting of a high-voltage-modulated photo-conductive-antenna was used, and an electro-optical-sampling method with a ZnTe nonlinear crystal was used to detect THz waves. THz waves were focused through a TPX (polymethylpentene) THz lens with a size of few millimetres, and samples were mounted at the focal point. Measurements of a sample and reference signal were averaged 10 times, and the entire system was purged with $N_2$ gas in a closed box, excluding the Ti:sapphire laser. For the nano-slot-antenna experiments, THz polarization perpendicular to the long axis of the slot was used. The measured transmittance, $T = \frac{I_{sam}(\omega)}{I_{ref}(\omega)} = \frac{(E_{sam}(\omega))^2}{(E_{ref}(\omega))^2}$, is obtained from the THz transmittance through a sample attached to an aperture, $I_{sam}(\omega) = (E_{sam}(\omega))^2$, divided by the THz transmittance through the void aperture, $I_{ref}(\omega) = (E_{ref}(\omega))^2$. $T$ for sugar sample was again normalized to the value through the empty antenna to compare the decreased transmittance ratio.

A far-field THz image in transmission was acquired by moving the sample stage at the THz focal point. The sample was attached to the square aperture having dimensions of 5 mm × 5 mm. A THz transmitted field was obtained for each pixel while the stage was moving with a step size of a 200 µm; thus, the entire image is 25 × 25 pixels. We averaged the THz time-domain waveforms three times and converted them to frequency-domain spectra using a fast Fourier transform (FFT) method, as in typical spectroscopy. Then, a contour colour image of the transmittance in space can be selected at certain frequency. The total data acquisition time was approximately 1.5 h for scanning a 5 mm × 5 mm area, and the entire system was also nitrogen-purged during the measurement.

## 2. The THz absorption coefficients and the complex refractive indexes

The samples consisting of monosaccharides, disaccharides, and polysaccharides were prepared in highly crystalline powder states at first. Then, the samples were thoroughly ground in a mortar and pressed in a pellet die under a pressure of approximately 2000 psi for 5 min. The average thicknesses of the samples are 650 µm on average, and the diameters are 8 mm, whereas the spot size of the THz beam is 2 mm, which is large enough to cover a homogeneous area of the pellets. Such pellet-type samples provide full spectral information including the inter-molecular vibrational modes of the targeted sugar molecules, which allows for full fingerprinting in the THz spectrum (0.5–2.0 THz)[24].

The refractive indexes for various sugars and sweeteners are extracted from the transmitted THz spectra in the range of 0.5–2.5 THz. The THz time-domain waveforms of a reference and sample were transformed to the frequency-domain using an FFT. The complex optical constants can be calculated with the relationship between the reference and sample waveforms in the frequency domain as follows:

$$E_{sam}(\omega) = E_{ref}(\omega) \cdot \exp\left(-\frac{d \cdot \alpha(\omega)}{2}\right) \cdot \exp\left(i \frac{2\pi}{\lambda} n(\omega) d\right), \tag{1}$$

where $E_{sam}(\omega)$ is the amplitude of the transmitted signal through the sample, and $E_{ref}(\omega)$ is the amplitude of the input signal through empty space occupied with the sample. $n(\omega)$ and $\alpha(\omega)$ are the real parts of the refractive index and absorption coefficient, respectively, and $d$ is the thickness of the sample. From the difference between the spectral amplitudes which passed through the sample and reference, the power absorption was extracted, which is related to the imaginary part of the refractive index, $\kappa(\omega)$. $n(\omega)$ is obtained from the phase difference between the two signals as

$$n(\omega) = 1 + \frac{\varphi_{ref} - \varphi_{sam}}{2\pi d} \lambda, \tag{2}$$

and the absorption coefficient,

$$\alpha(\omega) = -\frac{2}{d}\ln(T) = -\frac{2}{d}\ln\left(\left(\frac{E_{sam}(\omega)}{E_{ref}(\omega)}\right)^2\right) = \frac{4\pi k}{\lambda}, \tag{3}$$

where $\varphi_{ref}$ is the phase of the reference waveform, $\varphi_{sam}$ is the phase of the signal which passed through the sample, and $\lambda$ is the wavelength.

## 3. Nano antenna design and sugar molecule dropping

The nano-antenna arrays consist of two-dimensional void rectangular slots in a 150-nm-thick gold film on a double-side-polished 500-μm-thick silicon wafer fabricated by an e-beam lithography technique. Each rectangular slot antenna has a fundamental resonance frequency, $f_{res}$, which has the relation $f_{res} = \frac{c}{\lambda_{res}} = \frac{c}{\sqrt{2(n_{sub}^2+1)}\,l}$, where $l$ is the length of a rectangular slot, and $n_{sub}$ is the effective refractive index of the substrate[25]. We designed two different nano-slot-antenna arrays for specific sugars: the glucose antenna has a length of $l = 40$ μm (the targeted frequency, $f_{res} = 1.4$ THz, is for D-glucose absorption) and the fructose antenna has a length of $l = 35$ μm (the targeted frequency, $f_{res} = 1.7$ THz, is for fructose absorption) with the same width $w = 500$ nm for both. Both periods between antennas are 40 μm in the horizontal direction and 50 μm in the vertical direction. Further, these closely placed slot antennas numbering 1400 in total are sufficient to reduce the possible uncertainties from the random distribution of sugar molecules after the sugar solution is dropped.

Aqueous solutions were prepared by dissolving sugar molecules in distilled water at room temperature; then, a 0.4-μL solution was dropped onto a nano-slot-antenna-array sensing chip by a pipette. We dried the sensing chip at room temperature within a few minutes to remove the broad water absorption at a reliable THz frequency[26].

**Figure 1**

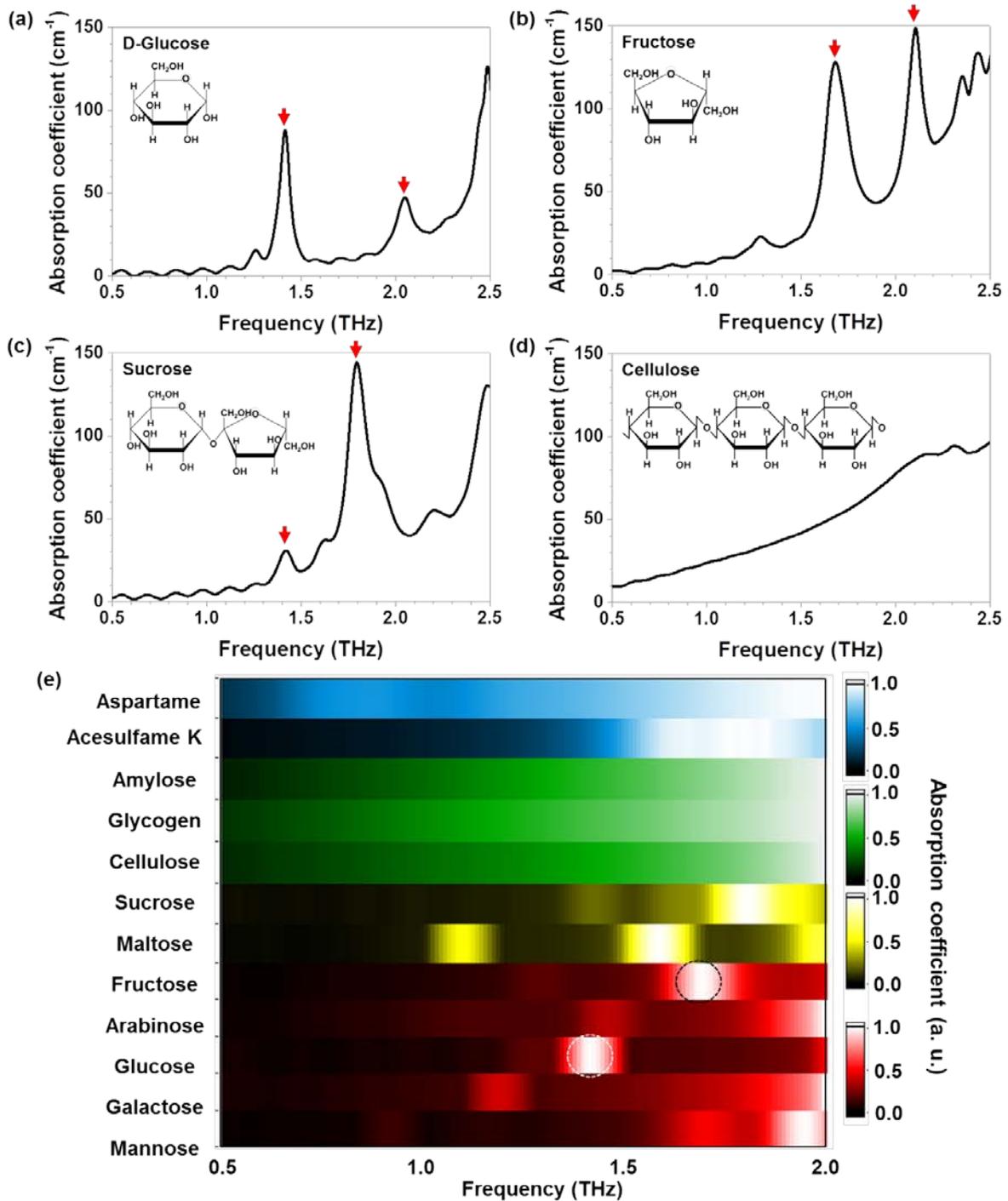

**Figure 1. Absorption coefficients for various sugars and sweeteners extracted from THz transmittance measurement.** Absorption coefficients at THz frequencies for (a) D-glucose ($C_6H_{12}O_6$), (b) fructose ($C_6H_{12}O_6$), (c) sucrose ($C_{12}H_{22}O_{11}$), and (d) cellulose (($C_6H_{10}O_5$)$_n$) pellets. The insets in (a)–(d) show the structural formulas of each saccharide. Distinguishable

absorption features appear at 1.4 THz for D-glucose, 1.7 THz and 2.1 THz for fructose, and 1.4 THz and 1.8 THz for sucrose, as marked by red arrows; however, cellulose has no special spectral features. (e) Colour contour plots of THz fingerprinting for ten different saccharides and two non-saccharide sweeteners. White and black dashed circles denote the specific frequencies of interest.

**Figure 2**

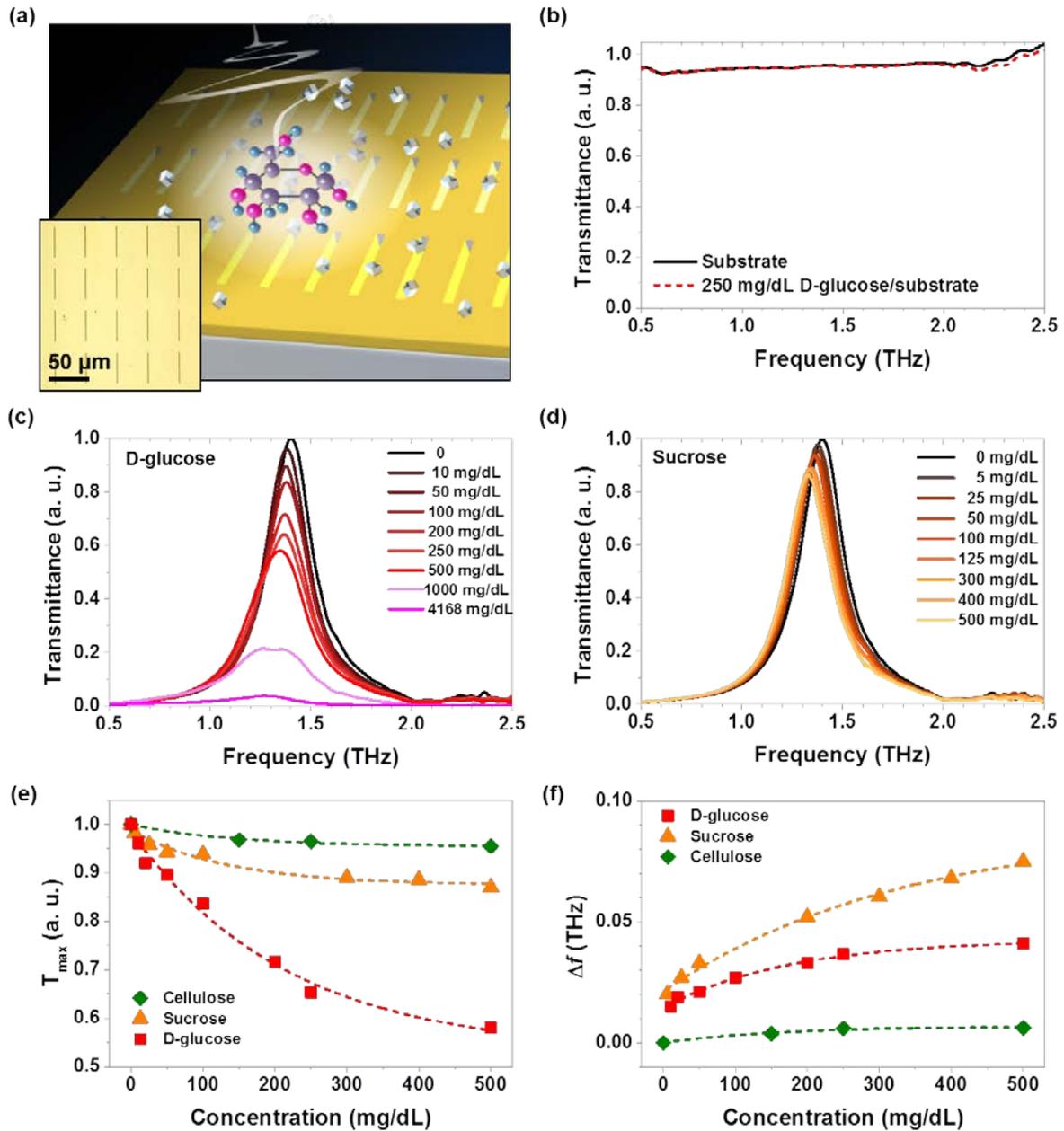

**Figure 2. The THz measurements for sugars with and without nano-antennas.** (a) Schematic of the THz detection of sugar molecules using a nano-antenna array-based sensing chip. Inset is a microscopic image of the nano-slot-antenna array. (b) Normalized THz spectra measured for a bare Si wafer used as a substrate and 250 mg/dL of glucose on the same Si substrate. (c) Normalized THz spectra measured with the glucose antenna for D-glucose and (d) sucrose molecules. (e) The changes in the maximum values of the normalized

transmittances are plotted for D-glucose, sucrose, and cellulose as a function of the molecular concentration level. (f) Frequency shifts at the maximum transmittance for three samples are plotted.

**Figure 3**

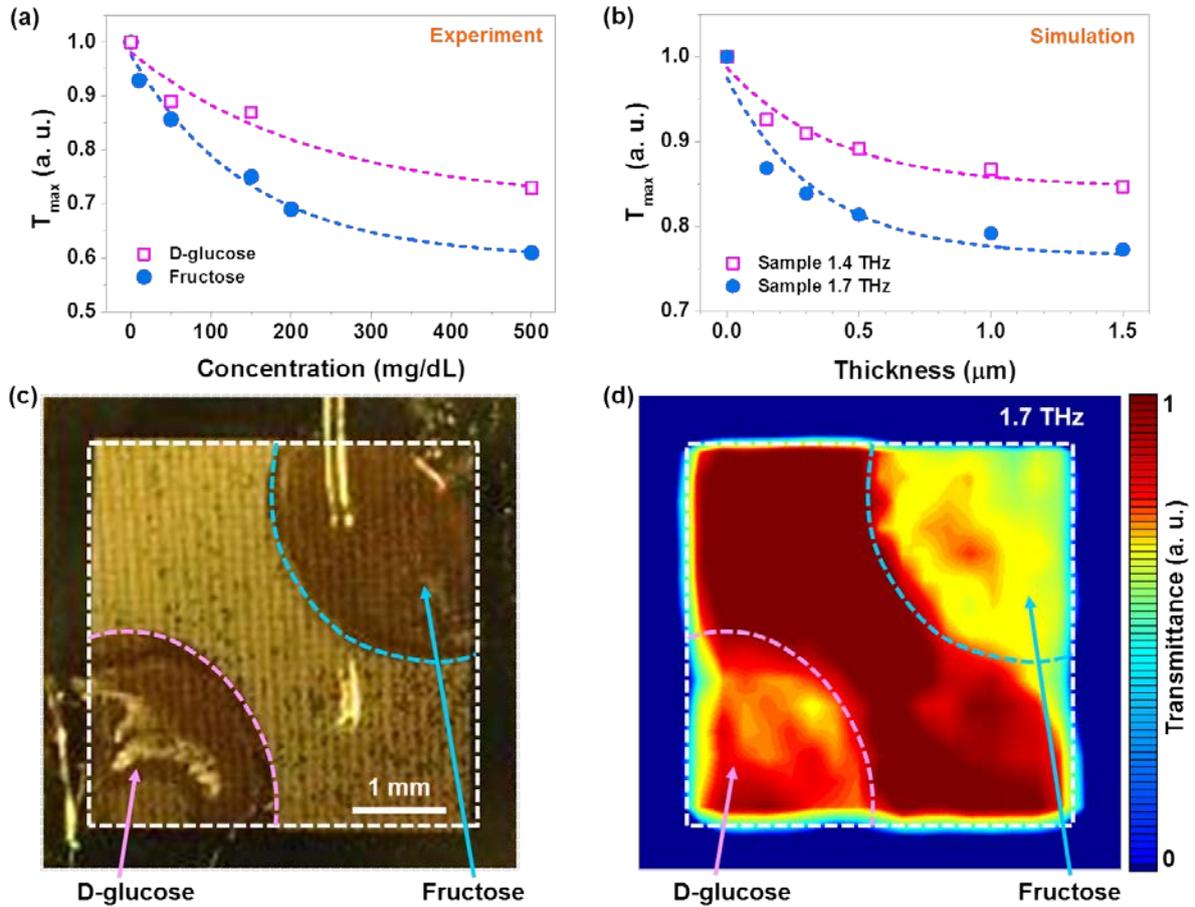

**Figure 3. A comparison with THz measurements and FDTD simulations, and THz images obtained with the fructose antenna.** (a) The changes in the maximum values of the normalized transmittances are plotted for fructose and D-glucose as a function of the molecular concentration level, measured using the fructose antenna. (b) Simulation results of cladding thickness-dependent maximum transmittances for two samples having absorption peaks at 1.7 THz and 1.4 THz are shown. The dashed lines are exponential fittings for all cases. (c) A photograph of the nano-antenna with 250 mg/dL of dropped fructose (upper-right corner) and D-glucose (lower-left corner) stains. (d) A normalized THz transmittance image through the fructose antenna with the two samples.

**Figure 4**

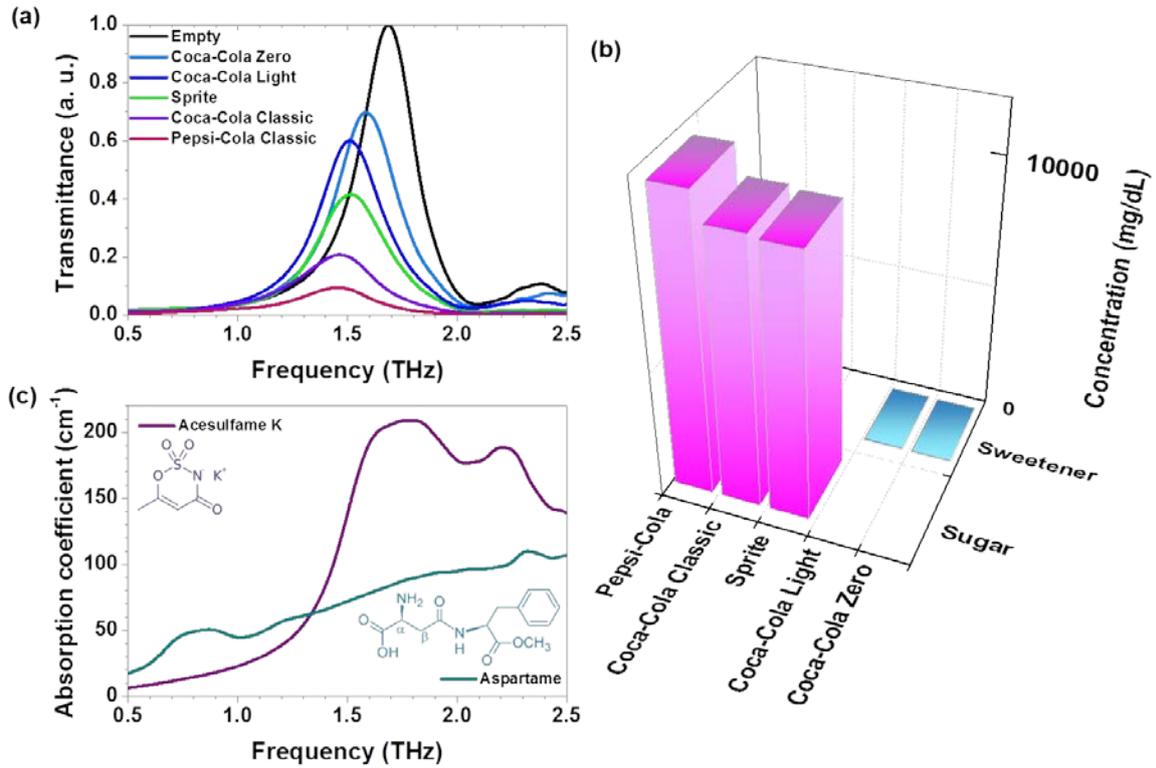

**Figure 4. The performance of the THz sugar antenna is verified with real market beverages containing acesulfame K, aspartame, sucrose, fructose, glucose, and so on.** (a) Normalized THz transmittances using a nano-antenna array with a fundamental resonance at 1.7 THz are shown for Coca-Cola Zero, Coca-Cola Light, Coca-Cola Classic, Pepsi-Cola, and Sprite. (b) The total concentrations of sugars and sweeteners in the real market beverages from the manufacturers. (c) Absorption coefficients for acesulfame K and aspartame. The insets show the structural formulas.

**Supporting Information**

**1. Crystallization effect**

Because absorption features appear clearly in crystalline states and ambiguously in amorphous states, the crystallized form is essential for efficient THz sensing. This fact has made it difficult to measure sugar molecules in an aqueous state after dissolution so far. In our experiment, sugar molecules were randomly distributed either on a metal side or at a slot-antenna gap after dropping and the dry process. Scanning electron microscopy (SEM) images, which were taken before (Fig. S1(a)) and after (Fig. S1(b)) THz measurements, show completely different forms of D-glucose molecules. A droplet of the aqueous sugar molecules experienced an amorphizing process, as shown in Fig. S1(a); however, well-crystallized D-glucose molecules inside the nano-slot antenna were formed by a possible THz-irradiation-induced heating effect. The THz-electric-field effect can be also considered to be responsible for the crystallization. It is known that packing of the same molecules with different orientations leads to the formation of a different type of crystal structure during the crystal-growth process[27]. By assisting with the nano-gap structures, the irradiated THz field can act as a monitoring tool for the molecular packing configuration and induce molecular polycrystalline packing at the same time. To the best of our knowledge, this is the first time that the THz-field-induced crystallization process due to the highly aligned electric-field effect has been reported. This crystallized form of sugar molecules inside the nanogap leads to an increased sensing efficiency, even at a very low molecular concentration because the quantitative detection ability indicated in the THz spectra highly relies on the polycrystalline state[28].

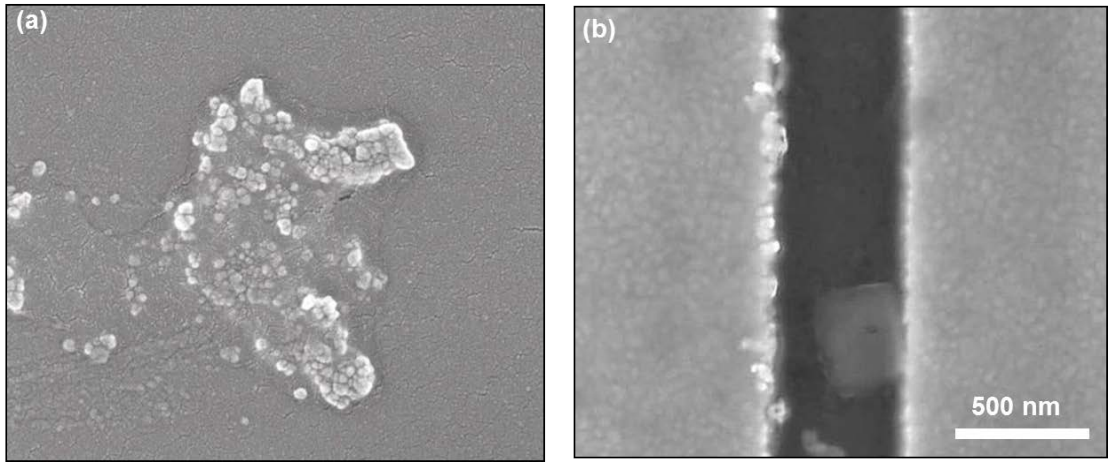

**Figure S1.** (a) A scanning electron microscopy (SEM) image of D-glucose molecules on a bare Si surface. (b) An SEM image of crystallized D-glucose molecules inside the nano-slot antennas after THz irradiation.

## 2. The complex refractive indexes for sugars and sweeteners

From the equation (3), the complex refractive indexes for the four sugars and two sweeteners were extracted (Fig. S2).

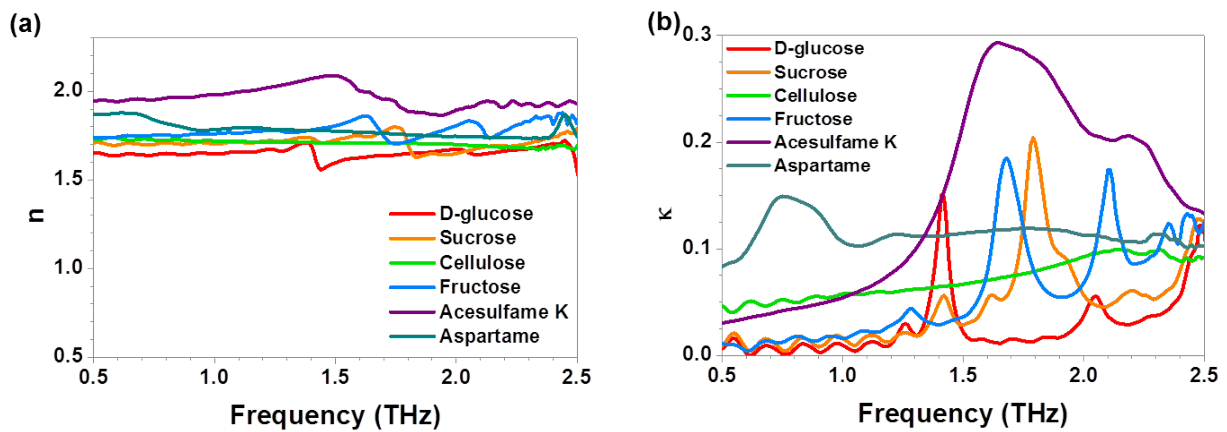

**Figure S2.** Complex refractive indexes for D-glucose, sucrose, cellulose, fructose, acesulfame K, and aspartame are extracted from the THz measurements. The (a) real part, $n$, and (b) imaginary part, $\kappa$, of the refractive index are plotted.

For example, the complex refractive indexes for several sugars and sweeteners extracted from the THz transmittance measurements using the specific sugar antenna are summarized in Table S1.

**Table S1. Complex refractive indexes for sugars**

|   | D-glucose (1.4 THz) | Sucrose (1.4 THz) | Cellulose (1.4 THz) | D-glucose (1.7 THz) | Fructose (1.7 THz) | Acesulfame K (1.4 THz) | Aspartame (1.4 THz) |
|---|---|---|---|---|---|---|---|
| $n$ | 1.65 | 1.74 | 1.71 | 1.64 | 1.76 | 1.96 | 1.76 |
| $\kappa$ | 0.15 | 0.05 | 0.065 | 0.015 | 0.18 | 0.14 | 0.11 |

$n$: real part of the refractive index

$\kappa$: imaginary part of the refractive index

## 3. FDTD simulations

The simulations were performed with a nano-antenna which has a fundamental resonance frequency at 1.7 THz, and we assumed that a higher molecular concentration yields thicker cladding deposition around the nano-antenna. In the FDTD calculations, the deposition is simplified as the formation of a uniform cladding on the nano-antenna such that the thickness of the cladding is assumed to be proportional to the molecular concentration which varies in the range of 0–1.5 µm. To mimic molecular-selective detection, we calculated the changes in the THz field transmittances through exemplary nano-antennas supporting the fundamental resonance at 1.7 THz affected by the cladding of two molecular samples possessing different absorption resonances at 1.7 THz and 1.4 THz.

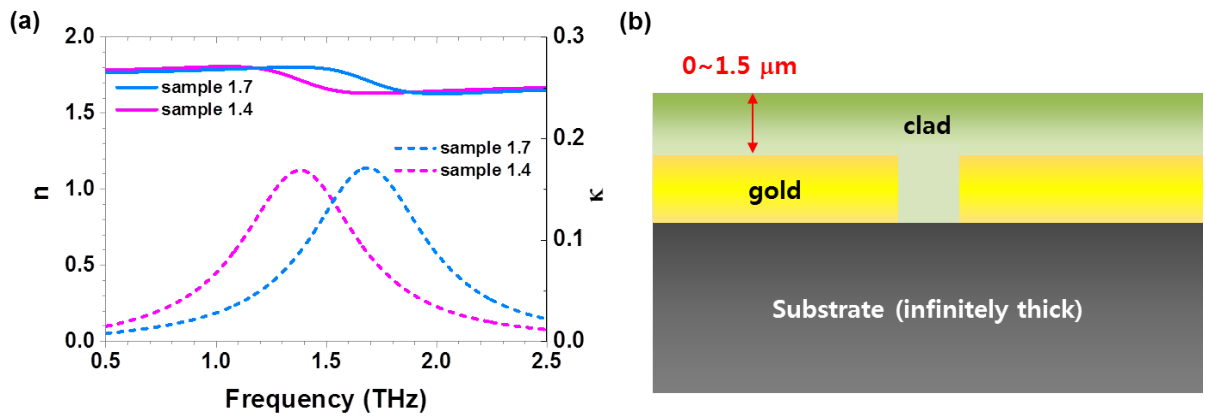

**Figure S3.** (a) Complex refractive indexes for two samples used in the FDTD simulations (*n* is plotted on the left axis, and $\kappa$ is plotted on the right axis). (b) Conceptual schematic for the simulation.

**Author Contributions**

D.-K.L. carried out the most of THz experiments with H.-S.K. and M.S., and prepared the samples with J.-S.L.. J.-H.K. calculated spectrums using FDTD method and carried out data analysis with Q.P.. C.K., J.H.K., T.L., and J.-H.S. contributed to analyze experimental data and prepared figures. M.S. developed the concept and wrote a paper. All authors discussed the results, and commented on and edited the manuscript.

**Acknowledgements**

This work was partially supported by the Centre for Advanced Meta-Materials funded by the Ministry of Science, ICT and Future Planning as a Global Frontier Project (Project No. 2014063727), and KIST intramural funding (2E25382 and 2E25723).

**Additional Information**

The authors declare no competing financial interests.